\documentclass[10pt,twocolumn]{revtex4}
\usepackage{graphicx,mathtools,amssymb}

\newcommand{\BE}{\begin{equation}}
\newcommand{\EE}{\end{equation}}
\newcommand{\BA}{\begin{eqnarray}}
\newcommand{\EA}{\end{eqnarray}}
\newcommand{\bx}{{\bf x}}
\newcommand{\bv}{{\bf v}}
\newcommand{\mP}{\mathbf{P}}
\newcommand{\mA}{\mathbf{A}}
\newcommand{\mQ}{\mathbf{Q}}

\begin{document}
\title{Lagrangian Flow Network approach to an open flow model}
\author{Enrico Ser-Giacomi$^1$, V\'{\i}ctor Rodr\'{\i}guez-M\'{e}ndez$^2$,
Crist\'{o}bal L\'{o}pez$^2$, Emilio
Hern\'{a}ndez-Garc\'{\i}a$^2$}
\affiliation{$^1$\'{E}cole Normale Sup\'{e}rieure, PSL Research
University, CNRS, Inserm, Institut de Biologie de l'\'{E}cole
Normale Sup\'{e}rieure (IBENS), F-75005 Paris, France \\
$^2$ IFISC (CSIC-UIB), Campus Universitat de les Illes Balears,
E-07122 Palma de Mallorca, Spain}
\begin{abstract}
Concepts and tools from network theory, the so-called
Lagrangian Flow Network framework, have been successfully used
to obtain a coarse-grained description of transport by closed
fluid flows. Here we explore the application of this
methodology to open chaotic flows, and check it with numerical
results for a model open flow, namely a jet with a localized
wave perturbation. We find that network nodes with high values
of out-degree and of finite-time entropy in the forward-in-time
direction identify the location of the chaotic saddle and its
stable manifold, whereas nodes with high in-degree and
backwards finite-time entropy highlight the location of the
saddle and its unstable manifold. The cyclic clustering
coefficient, associated to the presence of periodic orbits,
takes non-vanishing values at the location of the saddle
itself.
\end{abstract} 
\maketitle
\section{Introduction}
\label{sec:intro}

The use of simple kinematic flows to study chaotic transport
has allowed a deeper understanding  of its theoretical aspects
and its laboratory and environmental applications
\cite{Ottino1989,Wiggins1992}. In the particular context of
oceanic processes these chaotic models, complemented with tools
and methods of nonlinear dynamical systems
\cite{Tel2006,Cencini2010,mancho2006tutorial}, provided
advances in the study of ocean transport \cite{cencini1999},
marine particle dispersion
\cite{bracco2000,lacasce2008statistics}, the distribution of
marine organisms
\cite{martin2003,sandulescu2006,sandulescu2007,sandulescu2008}
and the dynamics of coherent structures
\cite{dovidio2004mixing,peacock2010introduction,bettencourt2012oceanic}.

More recently, new tools coming from the theory of Complex
Networks are complementing and extending the above results. The
powerful framework of network theory has become a standard
toolbox in many scientific fields ranging from social science
to climate \cite{BarratBook2008,Newman2010,donges2009a}. In the
context of fluid dynamics, Lagrangian Flow Networks (LFNs)
\cite{sergiacomi2015flow,sergiacomi2015most,sergiacomi2015dominant,lindner2017spatio,rodriguez2017clustering}
have been introduced as a coarse-grained representation of
transport in which small regions in the fluid domain are
interpreted as nodes of a network, and the transfer of mass
from one of these regions to another defines weighted links
among them. They are based on the concept of transport
operators (also called transfer or mapping operators; in fact
they are the Perron-Frobenius operators of the transport
dynamics)
\cite{froyland2003detecting,Froyland2005,froyland2007detection,singh2008mapping,dellnitz2009seasonal,Froyland2012Rossi,Tallapragada2013,Bolltbook2013}.

The LFN methodology has been applied in previous works to
closed chaotic flows, which are characterized by bounded
chaotic trajectories of fluid parcels. Typical properties of
the flow are explained in terms of networks measures: mixing
and dispersion are related with the degree and related
quantities \cite{sergiacomi2015flow}, betweenness centrality
highlights preferred transit nodes connecting distant regions
\cite{sergiacomi2015most}, closeness and eigenvector centrality
distinguish regions dominated by laminar or by strong mixing,
and identify structures related to invariant manifolds
\cite{lindner2017spatio}, network communities identify coherent
fluid regions \cite{sergiacomi2015flow,rossi2014hydrodynamic},
and so on. Contrarily to these, open flows are characterized by
the escape of fluid particles from the domain of interest, and
chaoticity is restricted to a subregion from which fluid
particles are continuously escaping. Their behavior is thus
very different and we present in this paper a first study of
the specificities of LFN built for open flows. A full
characterization of these goes through determining escape rate
and distribution, constructing non-attracting chaotic sets and
their invariant measures and dynamical invariants
\cite{Lai2011}. The objective of this paper is expressing some
of these quantities in the language of networks and illustrate
them with a simple model system.

The outline of the paper is as follows: in the next section we
review general open flow properties, then in Sect.
\ref{sec:network} we discuss them in terms of network
measurements. In Sec. \ref{sec:numerical} we check our results
with a simple flow model. In Sec. \ref{sec:conclusions} we
present our conclusions.

\section{Chaotic open flows}
\label{sec:open}

We summarize here the main properties of open chaotic flows,
stressing differences with closed ones. The Lagrangian
description of transport by a flow is characterized by the
equations of motion of a fluid particle in the velocity field
$\bv$
\BE
\dot\bx(t) = \bv(\bx(t),t).  \label{eqmotion}
\EE
By integrating this equation for different initial conditions
the {\sl flow map} $\Phi_{t_0}^\tau$ is obtained. It gives the
position at time $t_0+\tau$ of the fluid particle started at
$\bx_0$ at time $t_0$:
\BE
\bx(t_0+\tau)=\Phi_{t_0}^\tau (\bx_0) \ .
\label{map}
\EE
Evaluation at every initial condition inside a set $A$ defines
the action of the flow map on the fluid region,
$\Phi_{t_0}^\tau(A)$.

A distinctive local characteristic of the dynamical system
(\ref{eqmotion}) or (\ref{map}) is the Finite Time Lyapunov
Exponent (FTLE). It is defined as \cite{Ott1993,Shadden2005}
\BE
\lambda(\bx_0,t_0;\tau)=\frac{1}{2|\tau|}\log |\Lambda_{max}|
\label{FTLE}
\EE
with $\Lambda_{max}$ the largest eigenvalue of the right
Cauchy-Green strain tensor:
\BE
C(\bx_0,t_0,\tau) = \left(\nabla
\Phi_{t_0}^\tau(\bx_0)\right)^T \nabla \Phi_{t_0}^\tau(\bx_0) \
.
\EE
$\nabla \Phi_{t_0}^\tau(\bx_0)$ is the Jacobian matrix of the
flow map, and $M^T$ means the transpose of the matrix $M$. If
$\tau>0$ this is the forward FTLE. If instead trajectories are
computed backwards in time ($\tau<0$), then we obtain the
backwards FTLE field. The Lyapunov exponent characterizes the
typical rate of separation, averaged in an interval of time
$\tau$, of infinitesimally close initial conditions located
around $\bx_0$ at time $t_0$. In two-dimensional (2d) flows
there is a second eigenvalue of $C(\bx_0,t_0,\tau)$, which
defines a second Lyapunov exponent, say
$\lambda'(\bx_0,t_0;\tau)$, via a formula similar to Eq.
(\ref{FTLE}). In 2d incompressible flows, the case that will be
considered here, we have
$\lambda(\bx_0,t_0;\tau)=-\lambda'(\bx_0,t_0;\tau)$. The
dependence on $t_0$ is determined by the time dependence of
$\bv(\bx,t)$. For example, if the velocity field is
time-periodic of period $T$, $\bv(\bx,t)=\bv(\bx,t+T)$, the
same holds for the FTLE:
$\lambda(\bx_0,t_0;\tau)=\lambda(\bx_0,t_0+T;\tau)$.

Under standard conditions \cite{Tel2006,Ott1993}, as
$\tau\to\infty$ the FTLE approaches a constant value $\lambda$,
called \emph{the} Lyapunov exponent, at \emph{almost all}
points $\bx_0$ in an ergodic region, a positive value of this
quantity being a common indicator of chaotic behavior. Strong
inhomogeneities typically persist, however, in sets of
locations $\bx_0$ of zero Lebesgue measure. This dependence on
$\bx_0$, often of filamental aspect, becomes more evident at
intermediate $\tau$ and has been used to characterize important
transport structures (Lagrangian coherent structures)
\cite{haller2000lagrangian,shadden2005definition,haller2015lagrangian}.
In particular, for properly chosen values of $\tau>0$, the
forward FTLE, $\lambda(\bx_0,t_0;\tau)$, tends to take large
values for $\bx_0$ on stable manifolds of strong hyperbolic
trajectories or structures, whereas large values of the
backwards FTLE, $\lambda(\bx_0,t_0;-\tau)$, tend to highlight
the location of unstable manifolds
\cite{shadden2005definition}. Homoclinic and heteroclinic
connections and tangles are also identified.

An open flow is one in which fluid leaves the domain of
interest, say $D$ (we do not consider here the possibility of
fluid entering the system). The quantity
\BE
S(D,t_0;\tau) \equiv e^{-\tau\kappa(D,t_0;\tau)} \equiv
\frac{m\left(D\cap \Phi_{t_0+\tau}^{-\tau}(D)\right)}{m(D)}
\label{escape}
\EE
is the proportion of fluid initialized in $D$ at $t_0$ that
remains in $D$ after a time $\tau$. It defines the finite-time
escape rate $\kappa(D,t_0;\tau)$. In contrast with the FTLE,
this is not a local quantity defined at each point, but depends
on a whole region $D$. We will think here on $m(A)$ as the
Lebesgue measure -- area, volume, etc.-- of a set $A$, although
other measures, such as mass or heat content of the region,
could be considered. A probabilistic interpretation of Eq.
(\ref{escape}) is that it gives the probability for a particle
released at $t_0$ at a random position in $D$ to remain in $D$
after a time $\tau$. The probability density of escape times
$\tau$ is then $f(\tau)=-d S/d\tau$. If the flow simply sweeps
the fluid out of the region, as for example a simple constant
velocity field would do, no fluid remains in $D$ after some
time and then $\kappa(D,t_0;\tau\to\infty)=\infty$. But an
interesting situation happens when
$\kappa(D,t_0;\tau\to\infty)$ approaches a finite non-zero
limit, the asymptotic escape rate $\kappa$, meaning that there
is some fluid (in an exponentially decreasing amount)
circulating inside $D$ for arbitrarily long times. If for these
trajectories $\lambda$ is sufficiently large compared to
$\kappa$, fluid elements there are being stretched into thin
filaments which elongate faster than they can leave the system,
so that they pile up in a fractal manner. This reveals the
existence inside $D$ of the so-called chaotic saddle, which is
a non-attracting zero-measure fractal chaotic set traced by
fluid elements that never leave the system
\cite{Tel2006,Lai2011}. This object has stable and unstable
manifolds, which intersect at the saddle itself. In 2d flows,
the dimension of the saddle is given by $D_{saddle}= 2 (D_0
-1)$, where $D_0$ is the dimension of the stable and unstable
manifolds (they have the same dimension in incompressible
flows), given by $D_0 \approx 2 -\frac{\kappa}{\lambda}$ where
$\lambda$ is the positive average Lyapunov exponent of the
system in the mixing region. Typical trajectories close to the
stable manifold  approach and spend a long time close to the
saddle, undergoing transient chaotic behavior, to leave the
system along the unstable manifold after some time. The
trajectories starting rightly at the stable manifold approach
the saddle and move there chaotically, without never escaping
\cite{Tel2006,Lai2011}.

\section{The network approach}
\label{sec:network}

The network representation of fluid flow
\cite{sergiacomi2015flow,lindner2017spatio} uses the
set-oriented approach to transport
\cite{froyland2003detecting,Froyland2005,froyland2007detection,dellnitz2009seasonal},
and requires the discretization of the fluid domain $D$ in
small boxes, $\{B_i, i=1,2,...,N\}$, which are identified with
network nodes. Then, a directed link with a weight
$\mP(t_0,\tau)_{ij}$, the proportion of the fluid started in
$B_i$ which is found in $B_j$ after a time $\tau$, is assigned
to each pair of nodes $i,j$:
\BE
\mP(t_0,\tau)_{ij} = \frac{m\left(B_i \cap
\Phi_{t_0+\tau}^{-\tau}(B_j)\right)}{m(B_i)} \ . \label{PF}
\EE
$\mP(t_0,\tau)_{ij}$ is called the transfer or transport
matrix, and is a discrete approximation to the Perron-Frobenius
operator of the flow. $\mP(t_0,\tau)_{ij}$ can be interpreted
as the probability for a particle to reach the box $B_j$, under
the condition that it started from a uniformly random position
within box $B_i$. In the network approach, $\mP(t_0,\tau)$ is
the adjacency matrix of a weighted and directed network.

Numerical estimation of  $\mP(t_0,\tau)$ can be done by
releasing a large number $n_i$ of particles randomly placed in
box $B_i$, computing their trajectories for a time $\tau$, and
counting the number of particles arriving to each $B_j$
\BE
\mP(t_0,\tau)_{ij} \approx \frac{\textrm{\# of particles from
box $i$ to box $j$}}{n_i} \ . \label{PFparticles}
\EE
Note that this strategy immediately gives also the standard way
to compute $S(D,t_0;\tau)$ in Eq. (\ref{escape}): simply count
the fraction of the number of initially released particles
which still remain in $D$ after a time $\tau$. A standard
network-theory quantity, the out-strength of node $i$:
\BE
S_O(i)\equiv \sum_{j=1}^N \mP(t_0,\tau)_{ij} \equiv
e^{-\kappa_i(t_0,\tau) \tau}
\EE
gives the fraction of particles started in $i$ still in the
system, and defines a local finite-time escape rate
$\kappa_i(t_0;\tau)$ of box $B_i$ (we have not written explicitly the $t_0$
and $\tau$ dependence on $S_O(i)$). The global escape fraction (that defines the global
escape rate $\kappa$) is a kind of
weighted average of the $S_O(i)$'s: $S(D, t_0;\tau)=\sum_i
m(B_i) S_O(i) /m(D)$. In closed flows, $S_O(i)=1$ $\forall i$,
and the matrix $\mP(t_0,\tau)$ is row-stochastic, but for open
flows $S_O(i)<1$. One can define an alternative transfer matrix
$\mQ(t_0,\tau)$:
\BE
\mQ(t_0,\tau)_{ij} \equiv
  \begin{dcases}
  \frac{\mP(t_0,\tau)_{ij}}{S_O(i)}
  & \qquad \textrm{if }  \quad S_O(i) \ne 0 \quad ,  \nonumber \\
\qquad 0 & \qquad \textrm{if }  \quad S_O(i)  =  0 \quad .
   \end{dcases}
\EE
This matrix is now row-stochastic, i.e. $\sum_{j=1}^{N}
\mQ(t_0,\tau)_{ij}=1$. It represents the probability of
reaching $B_j$ conditioned to starting in $B_i$ \emph{and} to
still remaining in the system. Because of its restriction to
the non-escaped fluid, it represents effectively a closed-flow
network.

There is still another matrix which is used in the network
description of fluid transport, the binary version of $\mP$:
\BE
\mA(t_0,\tau)_{ij}=
   \begin{dcases}
        1  & \qquad \textrm{if } \qquad \mP(t_0,\tau)_{ij}>0
 \quad \textrm{and} \quad  i \neq j \quad , \nonumber \\
        0  & \qquad \textrm{if }  \qquad \mP(t_0,\tau)_{ij}=0
  \quad \textrm{or} \quad  i=j \ .
   \end{dcases}
 \label{binary}
\EE
Note that the same matrix results if using $\mQ$ instead of
$\mP$. Taken as an adjacency matrix, $\mA$ defines a directed
unweighted network in which the weight information in $\mP$ is
neglected. The out-degree of node $i$, i.e. the number of nodes
receiving fluid from $i$ can be computed as
\BE
K_O(i)\equiv \sum_{j=1}^N \mA(t_0,\tau)_{ij} \ .
\EE
Again we have not made explicit the dependence on $t_0$ and
$\tau$. The corresponding in-strength and in-degree can also be
defined:
\BA
S_I(i) &=& \sum_{j=1}^N \mP_{ji} \ ,\\
K_I(i) &=& \sum_{j=1}^N \mA_{ji} \ .
\EA

The paper \cite{sergiacomi2015flow} introduced a family of
network entropies $H_i^q(t_0,\tau)$, $q=0,1,2,...$ relating the
matrix $\mP$ to the statistics of FTLE in finite boxes for the
closed-flow case. The row-stochastic matrix $\mQ$ can be
interpreted formally as a transfer matrix defining a
closed-flow network. Then the definition and properties of the
entropies in \cite{sergiacomi2015flow} can be taken directly by
using $\mQ$ instead of the complete open-flow transfer matrix
$\mP$. In particular, in the case in which all boxes $\{B_i\}$
have the same measure (and then transfer matrices are computed
numerically by releasing the same number of particles in each
node) and $\tau>0$ the members $q=0$ and $q=1$ of the family
are defined by:
\BA
H_i^0(t_0,\tau) & \equiv & \frac{1}{\tau}\log K_O(i) \ ,\\
H_i^1(t_0,\tau) & \equiv & -\frac{1}{\tau} \sum_{j=1}^N
\mQ(t_0,\tau)_{ij} \log \mQ(t_0,\tau)_{ij} \ .
\EA
$H_i^1$ is the finite-time entropy of
\cite{froyland2012finite}.  Reference \cite{sergiacomi2015flow}
related $H_i^0$ and $H_i^1$ in the closed flow case to averages
over $\bx_0$ in the box $i$ of quantities related to the FTLE,
namely $e^{\tau H_i^0}=K_O(i) \approx  \left<
e^{\tau\lambda(\bx_0,t_0,\tau)}\right>_{B_i}$ and $H_i^1
\approx \left<\lambda(\bx_0,t_0,\tau)\right>_{B_i}$. Here these
expressions will be modified by the escape process, but the
heuristics used in \cite{sergiacomi2015flow} still suggests
that $K_O(i)$ and $H_i^1$ take high values in boxes $i$ inside
which $\lambda(\bx_0,t_0;\tau)$ is large, i.e. on the saddle
and on its stable manifold.

For $\tau<0$ the above quantities should be computed with the
time-reversed velocity field or, equivalently, by replacing the
matrix $\mQ(t_0,\tau)$ by the one giving the time-backwards
dynamics \cite{sergiacomi2015flow,froyland2012finite}:
\BE
\mQ^B(t_0,\tau)_{ij} \equiv \mQ(t_0,-\tau)_{ij} =
\frac{\mQ(t_0-\tau,\tau)_{ji}}{\sum_k \mQ(t_0-\tau,\tau)_{ki}}
\ .
\EE
Values of $H_i^1$ computed with this matrix should be large in
boxes where $\lambda(\bx_0,t_0;-\tau)$ is large, i.e. on the
saddle and its unstable manifold. Note also that the out-degree
values computed from $\mQ^B$ are related to the in-degree
values computed with $\mQ$.  As a consequence, we also expect
large values of $K_I(i)$ to be associated to the saddle and its
unstable manifold.

Another fundamental set of quantities in network theory are the
clustering coefficients. Generally speaking, the clustering
coefficient of a node measures the proportion of closed
triangles in the network having that node as a vertex.
\cite{newman2003structure,newman2009networks}. Different
clustering coefficients can be defined depending on the type of
network (weighted, directed, ...) and of the kind of triangles
one is interested in
\cite{saramaki2007generalizations,fagiolo2007clustering}. Of
interest here are \emph{cyclic triangles}. A cyclic triangle,
or 3-cycle motif, is one of the 3-node connected subgraphs
useful to characterize the local topology of networks
\cite{milo2002network}. It is a path in the network joining 3
nodes ($i$, $j$ and $k$) as $i \to j \to k \to i$. Given a node
$i$ with in-degree $K_I(i)$ out-degree $K_O(i)$ and with
$K_B(i)$ of these links being bidirectional ($K_B(i)=\sum_{j\ne
i} \mA_{ij} \mA_{ji}$), a \emph{cyclic clustering coefficient}
$C^c_i$ is defined as the ratio of all cyclic triangles
involving node $i$ present in the network, divided by all
possible cyclic triangles that could have been constructed with
these values of $K_I(i)$, $K_O(i)$ and $K_B(i)$. It can be
computed \cite{fagiolo2007clustering} from the diagonal
elements of the third power of the adjacency matrix $\mA$ :
\BE
C^c_i = \frac{\left(\mA(t_0,\tau)^3\right)_{ii}}{K_I(i)
K_O(i)-K_B(i)} \ . \label{cc}
\EE
$C_i^c$ takes values in $[0,1]$. Since it is constructed from
$\mA$ which neglects any weight information, the important
point is whether $C^c_i$ is zero or not at node $i$. If it is
non-vanishing then there is at least one directed triangle
involving $i$ in the network.

In Ref. \cite{rodriguez2017clustering} it was shown that, under
the standard approximation of Markovian dynamics (i.e.
$\mP(t_0,\tau_1+\tau_2)\approx
\mP(t_0,\tau_1)\mP(t_0+\tau_1,\tau_2)$)
\cite{dellnitz2009seasonal,Froyland2012Rossi,froyland2014how}),
for velocity fields either steady or periodic with period $T$,
and for values of $\tau$ multiple of $T$, $C^c_i$ is non-zero
at nodes containing the position at time $t_0$ of a periodic
trajectory of period $3\tau$. In open flows, periodic orbits
can only appear on the non-escaping set, i.e. the saddle. Thus,
we expect non-vanishing values of $C^c_i$ to identify the
saddle location. Generalized clustering coefficients involving
paths with more that 3-nodes can be considered, but we showed
in \cite{rodriguez2017clustering} that they lead to noisier
results.

In summary, our expectation on the properties of the
coarse-grained description of transport given by the Lagrangian
Flow Network methodology is that the saddle and its stable
manifold, where forward FTLE's take large values, are also
identified by nodes with high values of the out-degree $K_O(i)$
and of the forward finite-time entropy $H_i^1$. Analogously,
the saddle and its unstable manifold, associated to large
values of the backward FTLE, would be highlighted by high
values of the in-degree $K_I(i)$ and of the backwards
finite-time entropy. Finally, non-vanishing clustering
coefficient values are to be found at the saddle. In next
Section we check numerically the validity of these expectations
for a particular example of open flow.

\section{Numerical results for an example open flow}
\label{sec:numerical}

\subsection{A perturbed jet as an example of open flow}
\label{subsec:jet}

We use a model flow introduced in
\cite{hernandez2004sustained}, in a plankton ecology context,
to model an oceanic jet perturbed by a localized wave-like
feature. We use it because it is particularly simple, but at
the same time it has non-ideal features such as the very slow
velocity in some regions which makes non-exact some of the
hypothesis used. These hypotheses are mainly the supposition of
hyperbolic behavior, and the assumption that $\tau$ is large
enough and the fluid boxes small enough to guarantee that the
image of each box after a time $\tau$ is a thin and long
filament \cite{sergiacomi2015flow}. The hypotheses are
reasonably fulfilled in the central ($|y| \lesssim l$) region
of the jet, but they are clearly non correct in the slow
regions outside it. Despite this non-ideality we see that our
expectations on the meaning of the different network
quantifiers are confirmed.

The velocity field $\bv=(v_x,v_y)$ is two-dimensional and
incompressible, and is written in terms of a streamfunction
$\Psi(x,y,t)$:
\BA
v_x &=& \frac{\partial\Psi}{\partial y}  \nonumber \\
v_y &=& -\frac{\partial\Psi}{\partial x} \ . \label{hamilton}
\EA
with
\BE
\Psi=\Psi_0 \tanh\left( \frac{y}{l} \right)+\mu
 \exp\left(-\frac{x^2+y^2}{2\sigma^2}\right) \cos\left( k (y-vt) \right)\ .
\label{streamfunction}
\EE

The first term is the main jet, of width approximately $l$,
flowing towards the positive $x$ direction with maximum
velocity $\Psi_0/l$ at its center. The wave-like perturbation
(the region of chaoticity), of strength $\mu$, is represented
by the second term. It is localized in a region of size
$\sigma$ around the point $(x,y)=(0,0)$, and the wavenumber and
phase velocity (towards the positive $y$ direction) are $k$ and
$v$, respectively. The complete velocity field is time-periodic
with period $T=2\pi/kv$.

Equations (\ref{hamilton}) and (\ref{streamfunction}) define a
time-periodic Hamiltonian dynamical system. This type of system
typically develops chaotic regions when increasing the strength
of the perturbation, $\mu$. But fluid leaves the region $D$, so
that we have the situation of chaotic scattering: particles
enter $D$ from the left, following essentially straight
trajectories, experience transient chaos when reaching the wave
region, and finally they leave the system. For $\mu$ large
enough, recirculation gives birth to a chaotic saddle in $D$.
We take $l=1$, $\Psi_0=2$, $\sigma=2$, $\mu=3$, $k=1$, and
$v=1$, giving a flow period $T=2\pi/kv=2\pi$. Our domain of
interest will be $D=\{(x,y) ~|~ -9 \le x \le 9, -5 \le y \le
5\}$, from which we monitor the particle escape.

\begin{figure}
\begin{center}
\includegraphics[width=\columnwidth]{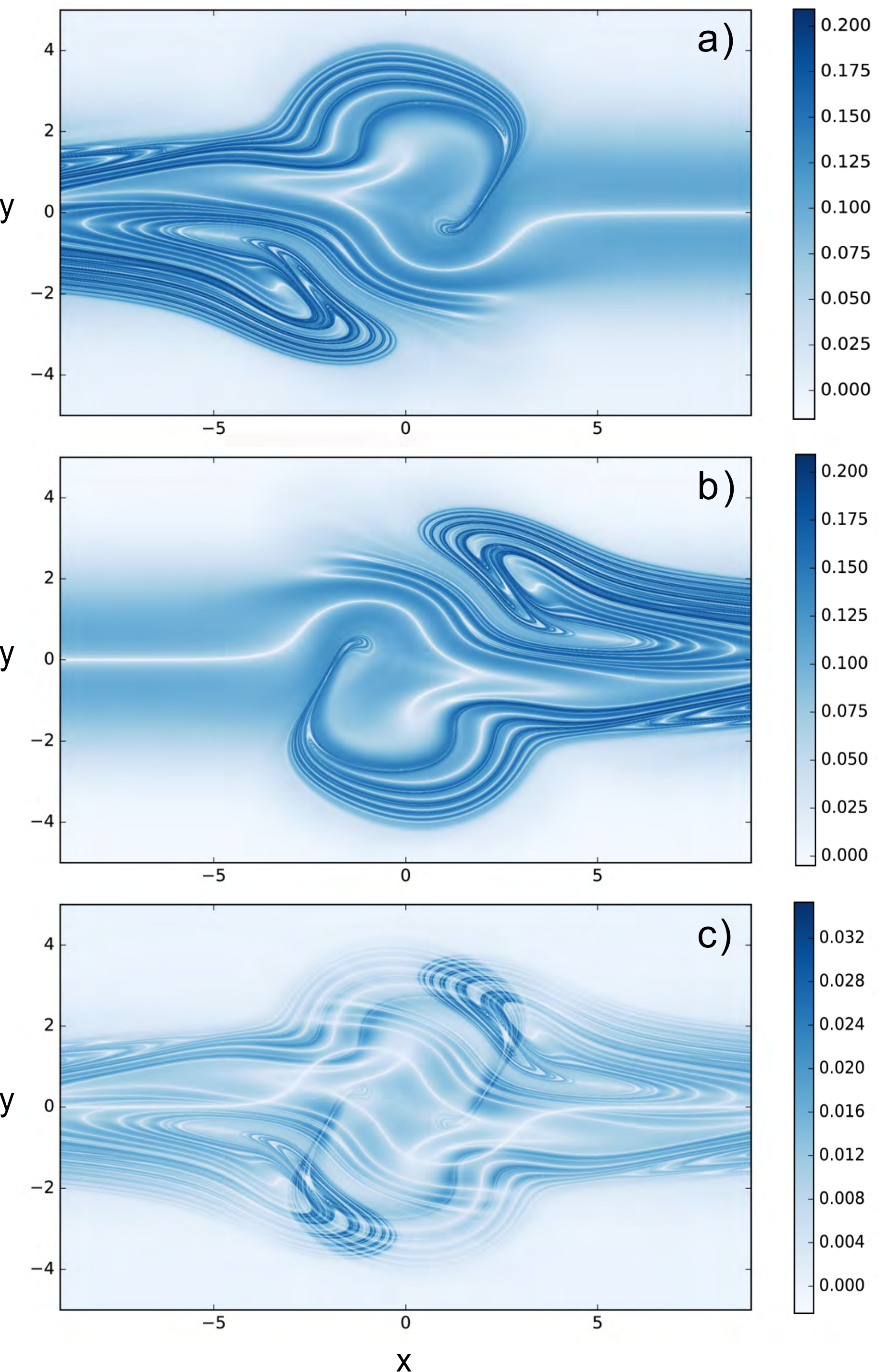}
\end{center}
\caption{Values of the FTLE at each initial location $\bx_0$, for $t_0=0$ and $\tau=6T$. a) Forward integration
(i.e. $\lambda(\bx_0,t_0,\tau)$); large values occur at the stable manifold of the chaotic saddle. b) Backward integration
(i.e. $\lambda(\bx_0,t_0,-\tau$); large values occur at the unstable manifold of the chaotic saddle.
c) The product $\lambda(\bx_0,t_0,\tau)\lambda(\bx_0,t_0,-\tau)$, which is large on the chaotic saddle.}
\label{fig:FTLE}       
\end{figure}

Figure \ref{fig:FTLE} displays the values of the FTLE, for
$t_0=0$, $\tau=6T$, and $\bx_0$ on a grid of spacing $0.01
\times 0.01$. The computation has been done by following all
trajectories for the full interval $\tau$ without taking into
account whether they remain inside $D$ or rather they leave the
domain. As expected, despite the information on the escape is
not explicitly taken into account, large values of he FTLE
identify filamental structures that reveal the locations of the
stable (top) and unstable (middle) manifolds (compare with Fig.
1b of Ref. \cite{hernandez2004sustained}). Bottom panel
displays the product of forward and backwards FTLE,
$\lambda(\bx_0,t_0;\tau)\lambda(\bx_0,t_0;-\tau)$, which takes
large values at the intersection of the two manifolds, and then
reveals the location of the chaotic saddle. We note that, for
this particular flow, determining the saddle and their unstable
and stable manifolds by the standard method
\cite{Tel2006,Lai2011} of plotting the locations of the
nonescaping particles at the middle, final, and initial times
is rather inaccurate. The reason is the nearly vanishing
velocity field at points with $|y|/l$ not close to the center
of the jet ($y=0$), say $|y| \gtrsim 3$. Trajectories started
at these points will finally escape the system, but only after
unpractically long integration times $\tau$. The FTLE
computation, however, clearly distinguishes the saddle and
manifold regions because of their large finite-time stretching
effect on the fluid elements.

\subsection{Network construction and analysis}
\label{subsec:results}

To construct the flow network, we discretize $D$ into $N=180
\times 100=18000$ boxes of size $0.1\times 0.1$. We release
initially (taking $t_0=0$) 100 particles inside each box and
compute their final position after a time $\tau=6T=37.699$. By
counting the particles interchanged between each pair of boxes
we compute $\mP(t_0,\tau)$ and the associated matrices
$\mA(t_0)$ and $\mQ(t_0,\tau)$ (and $\mQ^B(t_0,\tau)$), from
which we calculate the different network-node properties
defined above. Because of the escaping particles, $S_O(i)<1$ in
most nodes. The exception are many nodes in $|y| \gtrsim 3$ for
which, as stated above, the velocities are so small that
particles remain essentially immobile. An estimation of the
escape rate excluding this region, and thus characteristic of
the saddle, is $\kappa(D,t_0;\tau)\approx 0.03 \pm 0.01$ so
that the residence time is of the order of 33.33, or
approximately $5~T$.

\begin{figure*}
\begin{center}
\includegraphics[width=1.\textwidth]{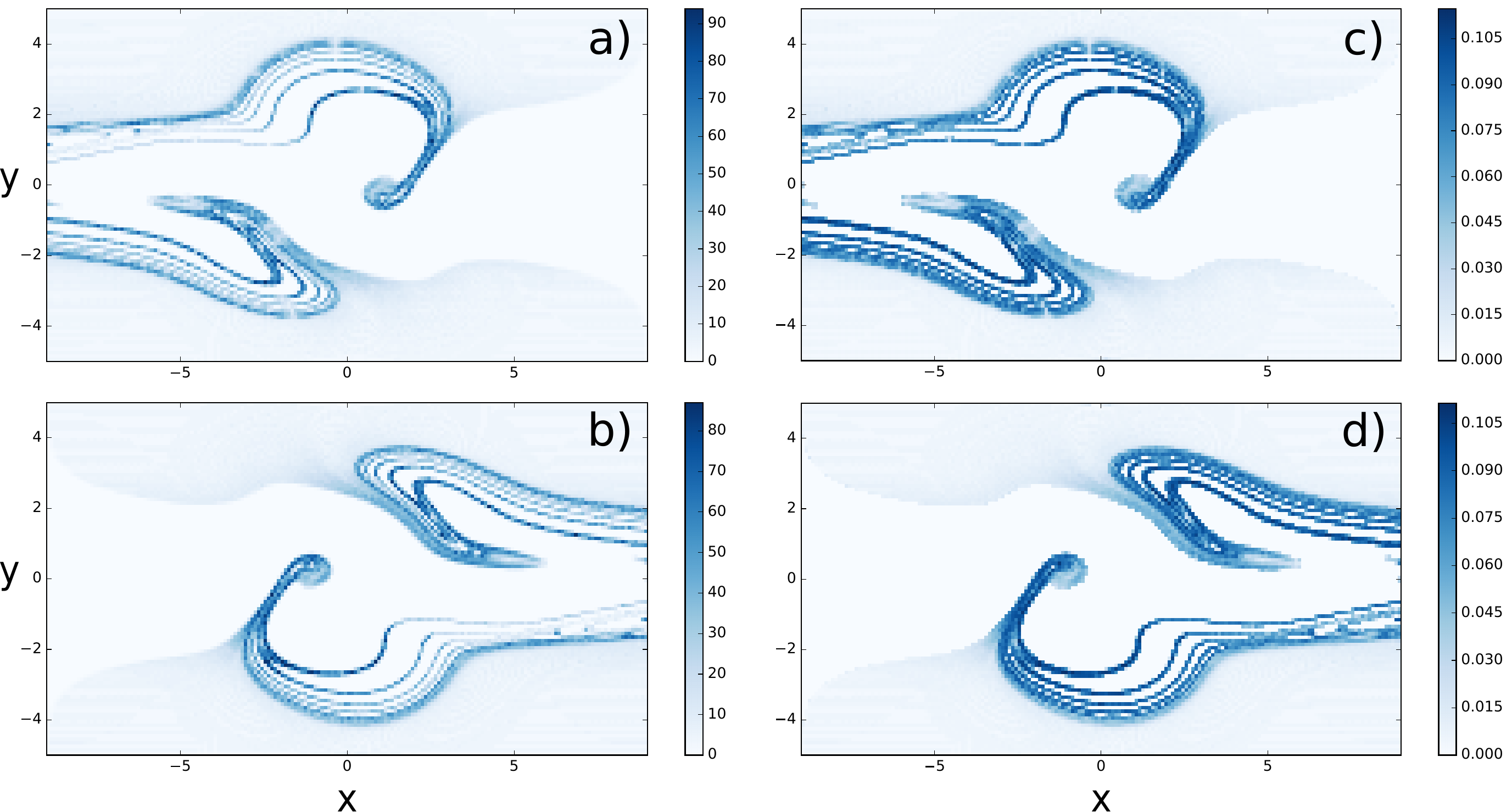}
\end{center}
\caption{Network quantifiers at the different nodes, giving a coarse-grained description of the dynamical
structures in the flow. a) Out-degree $K_O(i)$. b) In-degree $K_I(i)$. c)
Forward finite-time entropy $H_i^1(t_0,\tau)$. d) Backward finite-time entropy $H_i^1(t_0,-\tau)$. The upper panels
highlight the stable manifold of the saddle, and the lower ones its unstable manifold. }
\label{fig:degree_entropy}       
\end{figure*}

Figure \ref{fig:degree_entropy} shows the degrees $K_O(i)$ and
$K_I(i)$ (left) and the network entropies  $H_i^1(t_0,\tau)$
and $H_i^1(t_0,-\tau)$ (right). The figures confirm that high
values of these quantities identify the stable and unstable
manifolds of the non-escaping set, as revealed by the Lyapunov
fields in Fig. \ref{fig:FTLE}.

\begin{figure}
\begin{center}
\includegraphics[width=\columnwidth]{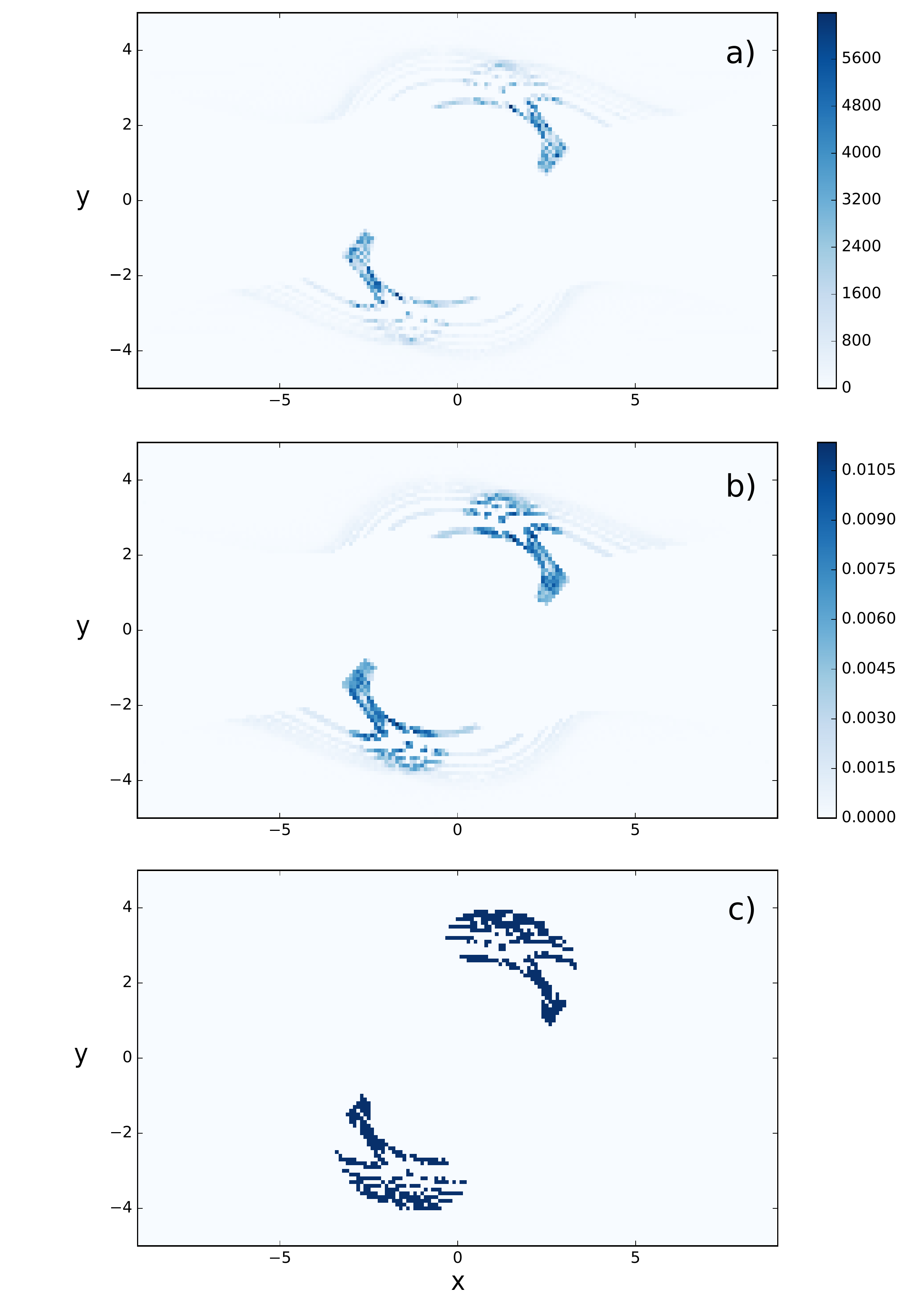}
\end{center}
\caption{Three network quantifiers giving a coarse-grained identification of the chaotic saddle.
a) Values of the degree product $K_O(i) K_I(i)$. b) Values of the  entropy product $H_i^1(t_0,\tau) H_i^1(t_0,-\tau)$.
c) Nodes with non-vanishing values of the cyclic clustering coefficient $C_i^c$.}
\label{fig:saddle}       
\end{figure}

Figure \ref{fig:saddle} shows three ways in which the network
approach locates a coarse-grained approximation to the saddle,
to be compared with the bottom panel of Fig. \ref{fig:FTLE}.
Top panel is the product $K_O(i)K_I(i)$, and middle panel is
the product $H_i^1(t_0,\tau)~H_i^1(t_0,-\tau)$. As expected,
these quantities take large values at the nodes that contain
the large values in Fig. \ref{fig:FTLE}c, i.e. nodes containing
pieces of the chaotic saddle. The bottom panel in Fig.
\ref{fig:saddle} shows the nodes with non-zero values of the
cyclic clustering coefficient $C^c_i$. They are
\cite{rodriguez2017clustering} the nodes containing at time
$t_0$ periodic trajectories of period $3\tau=18 T$, which can
only be present in this open system if embedded in the saddle.
Because of the large period $18 T$ involved, and to the finite
width of the node boxes, these nodes cover indeed most of the
chaotic saddle, as seen when comparing to panels a) and b) of
Fig. \ref{fig:saddle} and to Fig. \ref{fig:FTLE}c).

\section{Conclusions}
\label{sec:conclusions}

We have presented some numerical results, based on a simple
open flow model, on the description of dynamical properties of
advection by chaotic open flows within the framework of
Lagrangian Flow Networks. The network approach provides a
coarse-grained version of transport, and we expected that the
association of network nodes with high values of degree and
entropy to locations with high FTLE values, established
previously for closed flows, will remain valid at least
qualitatively for open flows. In particular, nodes with high
values of the out-degree $K_O(i)$ and of the forward
finite-time entropy $H_i^1(t_0,\tau)$ will give a
coarse-grained identification of the saddle and its stable
manifold, whereas nodes with large in-degree $K_I(i)$ and
backward finite-time entropy $H_i^1(t_0,-\tau)$ will highlight
the saddle and its unstable manifold. Non-vanishing values of
the cyclic clustering coefficient are to be found on periodic
orbits embedded on the saddle itself. We have numerically
checked these expectations and then confirmed that the
Lagrangian Flow Network methodology is a suitable framework to
characterize finite-time and coarse-grained view of transport
even for open flows with non-ideal characteristics.

As in the case of closed flows, we can not claim here that the
network approach is superior in all aspects to more specific
dynamical system tools. For example, Lyapunov-exponent
techniques for coherent structures were available
\cite{haller2000lagrangian,shadden2005definition,haller2015lagrangian}
before its reformulation in terms of networks
\cite{sergiacomi2015flow}, specific algorithms to find periodic
orbits in dynamical systems are well developed
\cite{Nayfeh1995}, as well as techniques to deal with open
systems \cite{Lai2011}. Usually the network approach requires
shorter trajectory integration, but this advantage is
compensated by the need to use many initial conditions to cover
the full domain. What is interesting in the network approach is
that it provides alternatives to all these sets of techniques
within a single framework, that the coarse-graining step
automatically tests for robustness to noise or diffusion, and
that it allows the use of techniques, such as community
detection or path-finding algorithms
\cite{sergiacomi2015flow,sergiacomi2015most,sergiacomi2015dominant},
beyond the scope of standard dynamical-systems approaches.
Future work will focus on the theoretical justification of our
heuristically derived and numerically confirmed relationships,
and in the development of additional network indicators more
specifically designed to describe open flows.

\section*{Acknowledgement}

We acknowledge financial support from grants LAOP,
CTM2015-66407-P (AEI/FEDER, EU) and ESOTECOS
FIS2015-63628-C2-1-R (AEI/FEDER, EU). ES-G received partial
support under the French program ``Investissements d'Avenir"
implemented by ANR (ANR-10-LABX-54 MEMOLIFE and
ANR-11-IDEX-0001-02 PSL Research University).

%


\begin{thebibliography}{45}

\bibitem{Ottino1989} J.~Ottino, \emph{The Kinematics of Mixing:
    Stretching, Chaos, and Transport}
  (Cambridge University Press, Cambridge, 1989)

\bibitem{Wiggins1992} S.~Wiggins, \emph{Chaotic Transport in
    Dynamical Systems} (Springer-Verlag, New
  York, 1992)

\bibitem{Tel2006} T.~T\'el, M.~Gruiz, \emph{Chaotic dynamics:
    {A}n introduction based on
  classical mechanics} (Cambridge Univ. Press, Cambridge, 2006)

\bibitem{Cencini2010} M.~Cencini, F.~Cecconi, A.~Vulpiani,
    \emph{Chaos: From simple models to complex
  systems} (World Scientific, Singapore, 2010)

\bibitem{mancho2006tutorial} A.M. Mancho, D.~Small, S.~Wiggins,
    Physics Reports \textbf{437}, 55 (2006)

\bibitem{cencini1999} M.~Cencini, G.~Lacorata, A.~Vulpiani,
    E.~Zambianchi, J. Phys. Oceanogr.
  \textbf{29(10)}, 2578 (1999)

\bibitem{bracco2000} A.~Bracco, J.H. LaCasce, A.~Provenzale, J.
    Phys. Oceanogr. \textbf{30(3)}, 461
  (2000)

\bibitem{lacasce2008statistics} J.~Lacasce, Progress in
    Oceanography \textbf{77}, 1 (2008)

\bibitem{martin2003} A.~Martin, Progress in Oceanography
    \textbf{57}, 125 (2003)

\bibitem{sandulescu2006} M.~Sandulescu,
    E.~Hern{\'a}ndez-Garc{\'\i}a, C.~L{\'o}pez, U.~Feudel,
    Tellus A
  \textbf{58}, 605 (2006)

\bibitem{sandulescu2007} M.~Sandulescu,
    E.~Hern{\'a}ndez-Garc{\'\i}a, C.~L{\'o}pez, U.~Feudel,
    Nonlinear
  Process. Geophys. \textbf{14}, 443 (2007)

\bibitem{sandulescu2008} M.~Sandulescu, C.~L{\'o}pez,
    E.~Hern{\'a}ndez-Garc{\'\i}a, U.~Feudel,
  Ecological Complexity \textbf{5}, 228 (2008)

\bibitem{dovidio2004mixing} F.~d'Ovidio, V.~Fern\'andez,
    E.~Hernandez-Garc\'ia, C.~L\'opez, Geophys. Res.
  Lett. \textbf{31}, L17203 (2004)

\bibitem{peacock2010introduction} T.~Peacock, J.~Dabiri, Chaos
    \textbf{20}, 017501 (~3) (2010)

\bibitem{bettencourt2012oceanic} J.~Bettencourt, C.~L\'{o}pez,
    E.~Hern\'{a}ndez-Garc\'{\i}a, Ocean Modell.
  \textbf{51}, 73 (2012)

\bibitem{BarratBook2008} A.~Barrat, M.~Barthelemy,
    A.~Vespignani, \emph{Dynamical processes on complex
  networks} (Cambridge Univ Press, Cambridge, 2008)

\bibitem{Newman2010} M.E.J. Newman, \emph{Networks: An
    Introduction.} (Oxford University Press,
  Oxford, 2010), ISBN 978-0-19-920665-0

\bibitem{donges2009a} J.F. Donges, Y.~Zou, N.~Marwan,
    J.~Kurths, The European Physical
  Journal-Special Topics \textbf{174}, 157 (2009)

\bibitem{sergiacomi2015flow} E.~Ser-Giacomi, V.~Rossi,
    C.~L\'opez, E.~Hern\'andez-Garc\'{i}a, Chaos
  \textbf{25}, 036404 (2015)

\bibitem{sergiacomi2015most} E.~Ser-Giacomi, R.~Vasile,
    E.~Hern\'andez-Garc\'{\i}a, C.~L\'opez, Physical
  Review E \textbf{92}, 012818 (2014)

\bibitem{sergiacomi2015dominant} E.~Ser-Giacomi, R.~Vasile,
    I.~Recuerda, E.~Hern\'andez-Garc\'{\i}a, C.~L\'opez,
  Chaos \textbf{25}, 087413 (2015)

\bibitem{lindner2017spatio} M.~Lindner, R.~Donner, Chaos
    \textbf{27}, 035806 (2017)

\bibitem{rodriguez2017clustering} V.~Rodr\'{\i}guez-M\'endez,
    E.~Ser-Giacomi, E.~Hern\'{a}ndez-Garc\'{\i}a, Chaos \textbf{27}, 035803
  (2017)

\bibitem{froyland2003detecting} G.~Froyland, M.~Dellnitz, SIAM
    Journal on Scientific Computing \textbf{24},
  1839 (2003)

\bibitem{Froyland2005} G.~Froyland, Physica D: Nonlinear
    Phenomena \textbf{200}, 205 (2005)

\bibitem{froyland2007detection} G.~Froyland, K.~Padberg, M.H.
    England, A.M. Treguier, Physical Review Letters
  \textbf{98}, 224503 (2007)

\bibitem{singh2008mapping} M.K. Singh, T.G. Kang, H.E.H.
    Meijer, P.D. Anderson, Microfluidics and
  Nanofluidics \textbf{5}, 313 (2008)

\bibitem{dellnitz2009seasonal} M.~Dellnitz, G.~Froyland,
    C.~Horenkamp, K.~Padberg-Gehle, A.~Sen~Gupta,
  Nonlinear Processes in Geophysics \textbf{16}, 655 (2009)

\bibitem{Froyland2012Rossi} G.~Froyland, C.~Horenkamp,
    V.~Rossi, N.~Santitissadeekorn, A.S. Gupta, Ocean
  Modelling \textbf{52}, 69 (2012)

\bibitem{Tallapragada2013} P.~Tallapragada, S.D. Ross,
    Communications in Nonlinear Science and Numerical
  Simulation \textbf{18}, 1106 (2013)

\bibitem{Bolltbook2013} E.M. Bollt, N.~Santitissadeekorn,
    \emph{Applied and Computational Measurable
  Dynamics} (SIAM, Philadelphia, 2013)

\bibitem{rossi2014hydrodynamic} V.~Rossi, E.~Ser-Giacomi,
    C.~L{\'o}pez, E.~Hern{\'a}ndez-Garc{\'\i}a,
  Geophysical Research Letters \textbf{41}, 2883 (2014)

\bibitem{Lai2011} Y.~Lai, T.~T\'el, \emph{Transient Chaos:
    Complex Dynamics on Finite-Time
  Scales} (Springer, 2011)

\bibitem{Ott1993} E.~Ott, \emph{Chaos in Dynamical Systems}
    (Cambridge Univ. Press, Cambridge
  (UK), 1993)

\bibitem{Shadden2005} S.C. Shadden, F.~Lekien, J.E. Marsden,
    Physica D \textbf{212}, 271 (2005)

\bibitem{haller2000lagrangian} G.~Haller, G.~Yuan, Physica D
    \textbf{147}, 352 (2000)

\bibitem{shadden2005definition} S.C. Shadden, F.~Lekien, J.E.
    Marsden, Physica D \textbf{212}, 271 (2005)

\bibitem{haller2015lagrangian} G.~Haller, Annual Review of
    Fluid Mechanics \textbf{47}, 137 (2015)

\bibitem{froyland2012finite} G.~Froyland, K.~Padberg-Gehle,
    Physica D: Nonlinear Phenomena \textbf{241},
  1612 (2012)

\bibitem{newman2003structure} M.E.J. Newman, SIAM Review
    \textbf{45}, 167 (2003)

\bibitem{newman2009networks} M.~Newman, \emph{Networks: An
    introduction} (Oxford University Press, 2009)

\bibitem{saramaki2007generalizations} J.~Saram\"aki,
    M.~Kivel\"a, J.P. Onnela, K.~Kaski, J.~Kert\'esz, Phys.
    Rev. E
  \textbf{75}, 027105 (2007)

\bibitem{fagiolo2007clustering} G.~Fagiolo, Phys. Rev. E
    \textbf{76}, 026107 (2007)

\bibitem{milo2002network} R. Milo, S. Shen-Orr, S.
    Itzkovitz, N. Kashtan, D. Chklovskii, U. Alon, Science \textbf{298},
    824 (2002)

\bibitem{froyland2014how} G.~Froyland, R.M. Stuart, E.~van
    Sebille, Chaos: An Interdisciplinary Journal
  of Nonlinear Science \textbf{24}, 033126 (2014)

\bibitem{hernandez2004sustained} E.~Hern\'andez-Garc\'{\i}a,
    C.~L\'opez, Ecological Complexity \textbf{1}, 253
  (2004)

\bibitem{Nayfeh1995} A.~H. Nayfeh and B. Balachandran,
    \emph{Applied Nonlinear Dynamics: Analytical, Computational and Experimental
Methods} (John Wiley, New York, 1995)

\end{thebibliography}

\end{document}